# Application of sodium-ion-based solid electrolyte in electrostatic tuning of carrier density in graphene

Jialin Zhao[1,2,3*], Meng Wang[1,2,3*], Xuefu Zhang[1,2], Yue Lv[1,2,3], Tianru Wu[1,2], Shan Qiao[1,2,3], Shufeng Song[4†], Bo Gao[1,2,3†]

[1]State Key Laboratory of Functional Materials for Informatics, Shanghai Institute of Microsystem and Information Technology, Chinese Academy of Sciences, 865 Changning Road, Shanghai 200050, China.

[2]CAS Center for Excellence in Superconducting Electronics (CENSE), Shanghai 200050, China

[3]University of Chinese Academy of Sciences, Beijing 100049, China

[4]College of Aerospace Engineering, Chongqing University, Chongqing 400044, P.R. China

**Abstract**

Using a solid electrolyte to tune the carrier density in thin-film materials is an emerging technique that has potential applications in both basic and applied research. Until now, only materials containing small ions, such as protons and lithium ions, have been used to demonstrate the gating effect. Here, we report the study of a lab-synthesised sodium-ion-based solid electrolyte, which shows a much stronger capability to tune the carrier density in graphene than previously reported lithium-ion-based solid electrolyte. Our findings may stimulate the search for solid electrolytes better suited for gating applications, taking benefit of many existing materials developed for battery research.

**Introduction**

Using an electric field to control electronic properties is an often-used method in materials science. It is also the key technique in the modern semiconducting industry. Traditionally, this technique relies on oxide-dielectric materials such as silicon dioxide. As an all-solid-state technique, it is easy to use and can be applied even at cryogenic temperatures. The main drawback of the oxide-dielectric-based technique is its weak capability to tune carrier density, which is usually limited to less than $1.0\times10^{13}$ cm$^{-2}$ because of the failure of the dielectric materials under high voltage[1]. Therefore, in the past, only the electronic transport of semiconducting materials could be effectively controlled with this technique. It is difficult to induce a considerable conductivity change in metals and superconductors. In the last decade, a new gating technique based on liquid electrolytes was developed. The liquid gating technique relies on the accumulation of ions near the interface between a liquid electrolyte and solid sample and can induce a carrier density change larger than $1.0\times10^{15}$ cm$^{-2}$, which is high enough to induce a considerable electrical conductivity change in gold films[2]. However, the coverage of liquid electrolytes on a sample surface makes this technique not compatible with many surface analysis techniques. The liquid nature of the electrolytes also impedes its application in making practical field-effect devices. To overcome the drawbacks of liquid electrolytes, solid electrolyte gating could be a solution. It has been demonstrated that protons embedded in microporous silicon dioxide could effectively control electronic transport in ITO[3]. However, the proton-based technique is quite sensitive to humidity in air, which limits its application[4]. Last year, a few groups successfully demonstrated that lithium-ion-based solid electrolytes can be used to tune the electronic properties of magnetic thin films[5], graphene[6] and iron-based superconductors[7]. These works have proven the feasibility of the solid electrolyte gating technique. Until now, only solid electrolytes based on small ions, such as protons or lithium ions, have been used. A natural question, then, is whether solid electrolytes based on larger ions can also demonstrate carrier density tuning capability. If the answer is yes, many solid electrolytes developed for battery research can be explored to improve the performance of the solid electrolyte gating technique. Another motivation to search for new solid electrolytes is to gain better control of the electro-chemical doping effect. Electrolyte gating is not a pure electrostatic doping process[8,9]. In both liquid and solid electrolyte gating processes, the mobile ions in the electrolytes may intercalate into the solid samples under high gate voltage, leading to the electro-chemical doping effect, which is not always a favourable factor. For example, lithium ions can intercalate into certain iron-superconductors to increase the superconducting transition temperature[7,10], but their intercalation into $NbSe_2$ leads to the opposite effect[11]. Because the intercalation of ions depends strongly on the ionic radius and the stability of ions in the lattice of the parent material, it is highly desirable to have more solid electrolytes based on different fast-moving ions to meet the growing needs to tune the electronic properties of a rich variety of thin-film materials. In this letter we report the use of a Nasicon-type sodium-ion-based solid electrolyte ($Na^+$-SE) to tune the carrier density in single-layer graphene. We show that it can induce a symmetric carrier density change in graphene, and its carrier density tuning capability is nearly one order of magnitude stronger than that of a previously reported Lithium-ion Conductive Glass-Ceramic (LICGC) solid electrolyte[6]. The double-layer capacitance of the solid electrolyte and the quantum capacitance of graphene were also extracted from the transport measurements data.

**Results**

**Gate voltage-dependent electronic transport measurements**

The $Na^+$-SE substrate is a white-looking material. The raw material is in cylinder form, as shown in the inset of Fig. 1a, with a diameter slightly less than 1 cm and a thickness of approximately 3.5 mm. The surface of the $Na^+$-SE substrates used in following experiments was hand-polished. Fig. 1a shows a scanning electron microscopy image of the polished surface of the electrolyte. The surface roughness is of approximately 30 nm measured from the atomic force microscopy image of a 5μm×5μm surface area.

Fig. 1b shows the electronic transport measurement results from a single-layer graphene sheet transferred onto a hand-polished $Na^+$-SE substrate. The resistance measurements were performed at 300 K in Helium atmosphere. Initially, the back-gate voltage ($V_g$) was swept towards the positive direction. The resistance peak appeared near 0.4 V. The voltage sweeping was interrupted at each selected positive voltage point to perform the Hall measurement. The discontinuity in the R vs. $V_g$ plot was caused by the delayed accumulation of sodium ions at the electrolyte/graphene interface during each Hall measurement[6]. After reaching the maximum gate voltage of 2.4 V, the gate voltage was swept back to zero in a continuous manner. There was a slight shift of the resistance peak position that was also caused by the delayed motion of sodium ions. We waited approximately eight hours at zero gate voltage to let the sample resistance relax to its initial value, as indicated by the dashed line in Fig. 1b. The back-gate voltage was then swept towards the negative direction. Hall measurements were also performed at selected negative gate voltage points.

**Dependence of carrier density and mobility on the gate voltage**

The ambipolar behaviour in the resistance measurement is preliminary evidence that the $Na^+$-SE substrate can effectively tune the Fermi level in graphene. To quantify the carrier tuning capability of the electrolyte, we deduced the carrier density and mobility from Hall measurement results. Fig. 2a and 2b show the Hall coefficient and the carrier density, respectively, measured at specific gate voltage points. The sign change confirms that the Fermi level was tuned across the Dirac point. On the positive gate voltage side, the electron density reached $5.0 \times 10^{14}$ $cm^{-2}$ at 2.4V; on the negative voltage side, the hole density reached $3.0 \times 10^{14}$ $cm^{-2}$ at −1.2 V. Such high carrier density tuning capability is much stronger than that reported for LICGC solid electrolyte[6]. Indeed most of our samples can reach an electron density larger than $3.0 \times 10^{14}$ $cm^{-2}$ at a gate voltage near 2.5 V. Such performance is truly comparable and even stronger than that of a few commonly used liquid electrolytes such as DEME-TFSI and ABIM-TFSI[12]. We noticed a drop in the hole density near −1.4 V. Such behaviour is common in liquid/solid electrolyte gating experiments. Because graphene is a chemically stable, it was probably solid electrolyte/surface contaminants that decomposed at high voltage and thus changed the electro-chemical doping of the graphene sheet. The details of the electro-chemical reaction at the graphene/electrolyte interface are still under investigation.

Fig. 2c and 2d show the hall mobility as a function of the gate voltage and the inverse Hall mobility as a function of the carrier density. Near zero gate voltage, the Hall mobility has a peak value of approximately 476 cm² V⁻¹ s⁻¹. This value is nearly one order of magnitude lower than that of our single-layer graphene sheet transferred onto other substrates such as SiO$_2$. Fig. 2c also shows that the Hall mobility decreases with increasing electron or hole density. In single-layer graphene, the main scattering mechanism includes Coulomb scattering by impurities, short-range scattering by defects and phonon scattering. The overall mobility can be defined using a Matthiessen's rule[13,14]: $\mu_{total}^{-1} = \mu_C^{-1} + \mu_{sr}^{-1} + \mu_{ph}^{-1}$. The phonon scattering sets the limit of the mobility on the order of $10^5$ cm² V⁻¹ s⁻¹, which is much larger than the measured Hall mobility. We can therefore neglect its contribution. The Coulomb scattering is independent of the carrier density: $\mu_C \sim$ constant. The short-range scattering is inversely proportional to the carrier density: $\mu_{sr}^{-1} \sim n$. Therefore the total mobility can be modelled using the following linear form: $\mu_{total}^{-1} = a + b \cdot n$. We plot the inverse Hall mobility as a function of the carrier density in Fig. 2d, which fits well with the above linear model. We found that on the electron side, $a_e = 2.5 \times 10^{-3}$ cm⁻² V s and $b_e = 3.8 \times 10^{-17}$ V s; while on the hole side, $a_h = 3.4 \times 10^{-3}$ cm⁻² V s and $b_h = 5.8 \times 10^{-17}$ V s. The fitting coefficient "a" represents the Coulomb scattering and the coefficient "b" represents the short-range scattering.

**Investigation of the double-layer capacitance at the**

## electrolyte/graphene interface and quantum capacitance of graphene

Tuning of the carrier density in graphene can be viewed as a process of charging capacitors in series. The inset of Fig. 3a shows the equivalent electric circuit diagram, in which $C_{DL}$ represents the capacitance of the electric double layer formed at the electrolyte/graphene interface, $C'_{DL}$ is the electric double-layer capacitance between the electrolyte and gold gate electrode, and $C_Q$ is the quantum capacitance of single-layer graphene. $C_{DL}$ is a constant value and is determined by the dielectric constant of the solid electrolyte and the radius of the counter-ions. To simplify the calculation, we set $C'_{DL}$ equal to $C_{DL}$, as our electrolyte substrate was adhered to the gold gate electrode using conductive glue. $C_Q$ is proportional to the density of states at the Fermi level and can be written as $C_Q = e^2 \cdot dn/dE_F$, where n is the carrier density and $E_F$ is the Fermi energy measured from the Dirac point. The total capacitance $C_{total}$ can be calculated using $C_{total}^{-1} = 2C_{DL}^{-1} + C_Q^{-1}$. Usually, the electric double-layer capacitance is much larger than the quantum capacitance; so the total capacitance can be roughly approximated by the quantum capacitance. The carrier density $n = 1/\pi \cdot (E_F/\hbar v_F)^2$ will then be primarily limited by $C_Q$ and $\sim(0.73\times10^{14}~\text{cm}^{-2}~\text{V}^{-2}) \cdot V_g'^2$, where $V_g'$ is the back-gate voltage measured from the Dirac point. Fig. 3a plot the absolute values of the carrier density as a function of $V_g'$. The blue line is the second-order polynomial fitting of the carrier density. We found a fitting parameter of ~$1.0\times10^{14}$ cm$^{-2}$ V$^{-2}$, which is close to the above theoretical calculation. To better fit the carrier density, the electric double-layer capacitance $C_{DL}$ needs to be taken into account. We model the system as three capacitors in series. The voltage drops on the electric double-layer capacitors and on the quantum capacitor of graphene adds up to the back-gate voltage: $V'_g = (2\ en\ /\ C_{DL}\ ) + (\hbar v_F\sqrt{\pi n}/e)$. The red line in Fig. 3a is the fitting of the carrier density using the equation above. We use the electric double-layer capacitance and the Fermi velocity as the fitting parameters. We found that $C_{DL}$ is approximately ~116 μF cm$^{-2}$, which is one order of magnitude larger than the value found for the ionic liquid electrolyte BMIM-PF$_6$[15]. We also deduced from the fitting a reduced Fermi velocity ~$0.43\times10^6$ m s$^{-1}$, which is smaller than the value ~ $0.85\times10^6$ m s$^{-1}$ expected for the situation of weak electron-electron interaction in graphene[16]. A similar low Fermi velocity was previously reported in graphene irradiated with Ar$^+$ ions[17]. The Fermi velocity in graphene follows $\hbar v_F = 3ta/2$, where a is the nearest carbon-carbon atomic distance and t is the nearest-neighbour hopping integral[18]. The impurities at the electrolyte/graphene interface may reduce the amplitude of the nearest-neighbour hopping of electrons in graphene, which leads to a decrease in the Fermi velocity.

After finding the double-layer capacitance from the fitting of the carrier density, we can deduce the quantum capacitance of the single-layer graphene sheet. The total capacitance is defined as $C_{total} = e\,dn/dV_g$, shown as the orange curve in Fig. 3b. The quantum capacitance can be calculated from the total capacitance using the fitted double-layer capacitance, as shown by the blue curve in Fig. 3b. Both curves show a typical "V" form. The smallest value of the quantum capacitance found near the charge neutrality point is ~ 15.6 μF cm$^{-2}$. The quantum capacitance is defined as $C_Q = (2e^2/\hbar v_F \sqrt{\pi}) \cdot (|n| + |n^\star|)^{1/2}$.[15] The non-zero value of the quantum capacitance is due to the additional carrier density n$^\star$ induced by impurities. We deduced n$^\star$ ~ 6×10$^{12}$ cm$^{-2}$, which is consistent with the carrier density deduced from Hall measurements near the charge neutrality point.

**Discussions**

We summarise the advantages of Na$^+$-SE over lithium-ion-based solid electrolytes such as previously reported LICGC. First, Na$^+$-SE provides a much stronger electrostatic carrier density tuning capability. For a single-layer graphene sheet, Na$^+$-SE can induce an average electron density change larger than 3.0×10$^{14}$ cm$^{-2}$. In some samples we even achieved an electron density change larger than 5.0×10$^{14}$ cm$^{-2}$. As a comparison, LICGC can only induce an electron density change less than[6] 1.0×10$^{14}$ cm$^{-2}$. The carrier density tuning capability of Na$^+$-SE is even better than most commonly used liquid electrolytes.

Second, the ion radii of Na$^+$ (102 pm) is much larger than that of Li$^+$ (76

pm)[19]. Intuitively speaking, it will be more difficult for sodium ions to intercalate into solid samples compared with lithium ions. This will be a favourable factor if one wants to reduce the influence of the electro-chemical doping effect. For example, it is known that $Na^+$ is chemically inert and does not adhere to specific locations on graphene[15]. Sodium is also difficult to intercalate into graphite[19]. Therefore sodium-ion-based solid electrolyte can be an ideal material to realise electrostatic doping in single/multi-layer graphene. As shown in Fig. 1b, the graphene sample can restore its initial resistance after the gate voltage sweeping, suggesting that electrostatic doping dominates the gating process. The symmetric carrier density tuning on both the electron and hole sides in the gate voltage window ranging from -1.2 V to 2.4 V also points to an electrostatic gating effect. Our results thus confirm that the $Na^+$-SE primarily tunes the carrier density in single-layer graphene in an electrostatic manner within the above mentioned gate voltage window.

Of course, the electro-chemical intercalation process is not solely determined by the ion radius. $Na^+$-SE may also induce electro-chemical doping in certain thin-film materials, which is also an attractive factor for future study because it provides us a method beyond the traditional material synthesis technique to induce exotic electron behaviour in sodium-containing materials. In conclusion, the successful demonstration of the carrier density tuning capability of $Na^+$-SE further extends the research of solid electrolyte gating technique, and it suggests that the existing solid electrolytes are indeed a treasure waiting to be explored.

**Methods**

$Na_{3.1}Zr_{1.95}Mg_{0.05}Si_2PO_{12}$ was synthesised through a solid-state reaction combined with mechano-chemical synthesis[20]. First, a $Zr_{1.95}Mg_{0.05}O_{3.95}$ solid solution was prepared through mechano-chemical reaction in a high-energy ball mill for 2 hours (SPEX SamplePrep 8000M Mixer). A mixture of $ZrO_2$ (Inframat Advanced Materials, ⩾ 99.9%) and MgO was milled by alternating 30 minutes of milling with 30 minutes in standby mode to avoid excessive heating. The solid electrolyte with the formula $Na_{3.1}Zr_{1.95}Mg_{0.05}Si_2PO_{12}$ was synthesised through a solid-state reaction by mixing stoichiometric amounts of $Na_2CO_3$ (Sigma-Aldrich, ⩾99.5%), $SiO_2$ (Sigma-Aldrich, ⩾ 99%), $NH_4H_2PO_4$ (Sigma-Aldrich, ⩾98%), and $Zr_{1.95}Mg_{0.05}O_{3.95}$, and ball-milling with zirconium oxide balls for 2 hours. The precursors were decomposed at 900 °C for 12 hours in alumina crucibles, with repeated ball-milling for 2 hours. The calcined powders were then cold pressed and sintered at 1260 °C for 16 hours covered with the raw powders to avoid sodium loss. The obtained bulk solid electrolytes were hand polished using aluminum oxide sandpaper (P1200). Instead of water, ethanol was used during polishing to prevent water damage.

Single-layer graphene samples were initially grown on copper substrates by the chemical vapour deposition (CVD) technique. A single-layer graphene sheet was transferred onto the solid electrolyte using a wet transfer method. A 250-nm-thick Poly(methyl methacrylate) (PMMA) layer was first spin-coated on top of the graphene. The graphene/PMMA layer was detached from the copper substrate using a 10 wt. % $FeCl_3$ solution. The floating graphene/PMMA layer was first transferred onto a $Si/SiO_2$ substrate and rinsed successively in deionised water and isopropanol (IPA). It was then transferred to the solid electrolyte substrate in IPA. The PMMA capping layer was removed by acetone. The samples were then annealed in a hydrogen–argon atmosphere for 7 hours. The electrical contacts to graphene were made by pressing indium pellets onto graphene. The solid electrolyte substrate was attached to a chip-carrier using conductive glue. The gate voltage was applied to the backside of the substrate using conductive glue.

The datasets generated during and/or analysed during the current study are available from the corresponding author on reasonable request.

**Acknowledgement**

We thank X.M. Xie for helpful discussions. We acknowledge the support from the National Natural Science Foundation of China under Grant No. 11374321 and U1632272; from the "Strategic Priority Research Program (B)" of the Chinese Academy of Sciences under Grant No. XDB 04010600; and from Helmholtz Association through the Virtual Institute for Topological Insulators (VITI).
**Author Information**

**Affiliations**

**State Key Laboratory of Functional Materials for Informatics, Shanghai Institute of Microsystem and Information Technology, Chinese Academy of Sciences, 865 Changning Road, Shanghai 200050, China.**

**CAS Center for Excellence in Superconducting Electronics (CENSE), Shanghai 200050, China**

**University of Chinese Academy of Sciences, Beijing 100049, China**

Jialin Zhao, Meng Wang, Yue Lv, Shan Qiao, Bo GAO

**State Key Laboratory of Functional Materials for Informatics, Shanghai Institute of Microsystem and Information Technology,**

Chinese Academy of Sciences, 865 Changning Road, Shanghai 200050, China.

CAS Center for Excellence in Superconducting Electronics (CENSE), Shanghai 200050, China

Xuefu Zhang, Tianru Wu,

College of Aerospace Engineering, Chongqing University, Chongqing 400044, P.R. China

Shufeng Song

## Contributions

J. Zhao, M. Wang and Y. Lv performed the experiments. J. Zhao, M. Wang contributed equally to this work. X. Zhang and T Wu provided single-layer graphene sheets. S Song provided the solid electrolyte. S Song, S. Qiao contributed to the discussions. S Song and B. Gao designed the experiment and wrote the manuscript.

## Competing interests

The authors declare no competing financial interests.

## Corresponding Authors

Correspondence to Shufeng Song (sfsong@cqu.edu.cn) and Bo GAO (bo_f_gao@mail.sim.ac.cn)

**Figure legends:**

**Figure 1**: (a) Scanning electron microscope image of the surface of a hand-polished Nasicon type solid electrolyte substrate. Inset: photo of a hand-polished $Na^+$-SE substrate. (b) Resistance variation of a single-layer graphene sheet with back gate voltage. The arrows in the figure indicate the direction of the voltage sweeping. The dashed line at zero gate voltage describes the evolution of the sample resistance during the relaxation of the accumulated sodium ions at the solid electrolyte/graphene interface.

**Figure 2**: The Hall coefficient (a), carrier density (b) and Hall mobility (c) of a single-layer graphene sheet as a function of the back-gate voltage. The measurements were performed at room temperature. (d) The inverse Hall mobility as a function of the carrier density. Solid lines are the linear fitting curves.

**Figure 3**: (a) The carrier density as a function of the modified gate voltage measured from the charge neutrality point. The blue line is the second order polynomial fitting curve assuming the quantum capacitance of graphene is much smaller than the double-layer capacitance. And the red line is the fitting curve taking the contribution of the electric double-layer capacitance into account. (b) The total capacitance and the quantum capacitance of the graphene sheet calculated using the fitted double-layer capacitance. Both curves show a typical "V"-form.

Fig.1

(a) 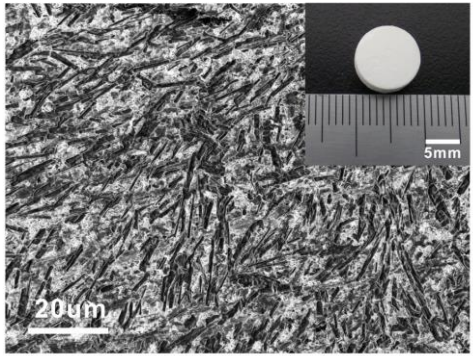 (b) 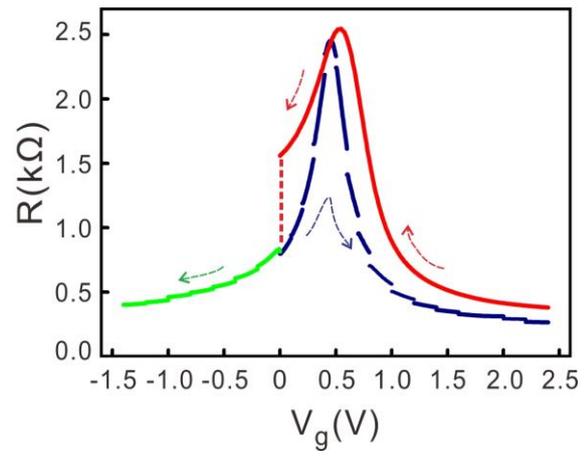

Fig.2

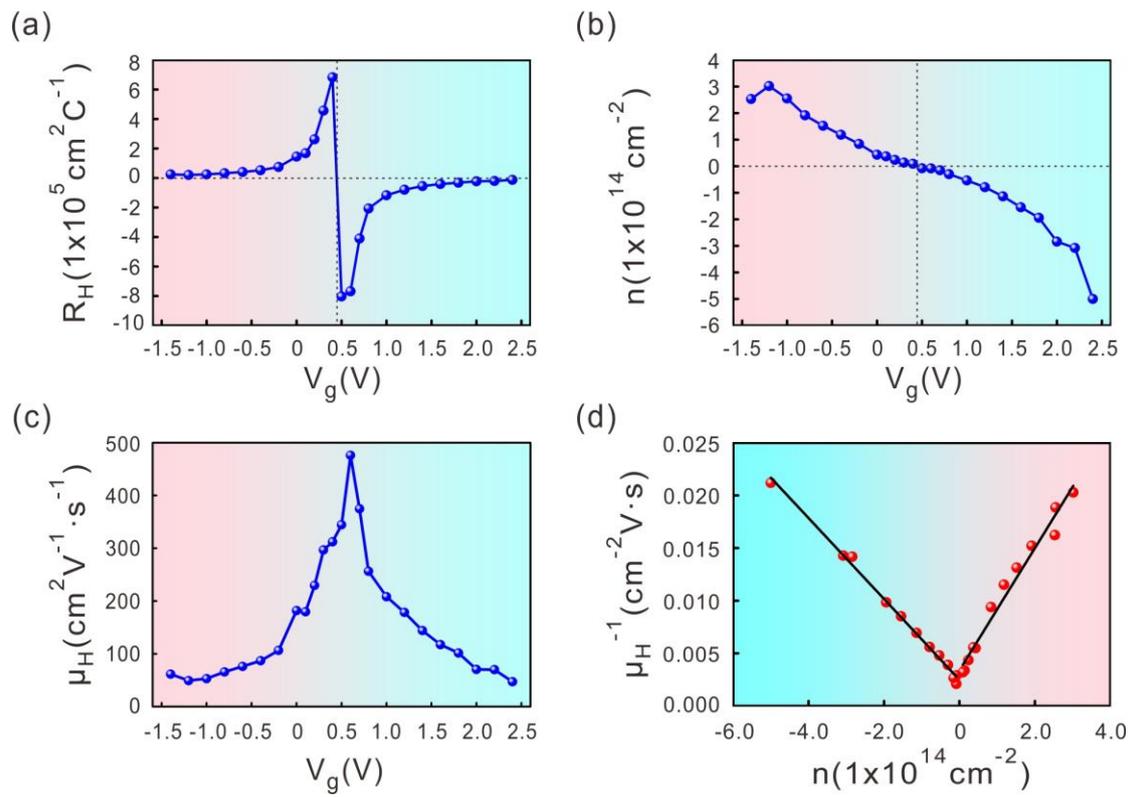

Fig.3

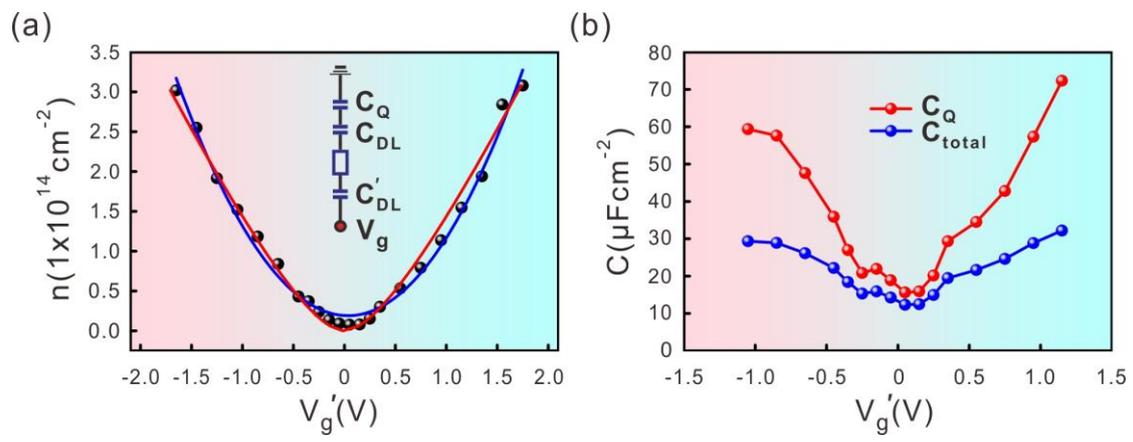